# GENERATIVE AI TOOLS IN ACADEMIC RESEARCH: APPLICATIONS AND IMPLICATIONS FOR QUALITATIVE AND QUANTITATIVE RESEARCH METHODOLOGIES

Mike Perkins [1*], Jasper Roe [2]

[1] British University Vietnam, Vietnam.
[2] James Cook University Singapore, Singapore
[*] Corresponding Author: Mike.p@buv.edu.vn

August 2024

## Abstract

This study examines the impact of Generative Artificial Intelligence (GenAI) on academic research, focusing on its application to qualitative and quantitative data analysis. As GenAI tools evolve rapidly, they offer new possibilities for enhancing research productivity and democratising complex analytical processes. However, their integration into academic practice raises significant questions regarding research integrity and security, authorship, and the changing nature of scholarly work. Through an examination of current capabilities and potential future applications, this study provides insights into how researchers may utilise GenAI tools responsibly and ethically.

We present case studies that demonstrate the application of GenAI in various research methodologies, discuss the challenges of replicability and consistency in AI-assisted research, and consider the ethical implications of increased AI integration in academia. This study explores both qualitative and quantitative applications of GenAI, highlighting tools for transcription, coding, thematic analysis, visual analytics, and statistical analysis. By addressing these issues, we aim to contribute to the ongoing discourse on the role of AI in shaping the future of academic research and provide guidance for researchers exploring the rapidly evolving landscape of AI-assisted research tools and research.

# Introduction

The development of Generative Artificial Intelligence (GenAI) tools has introduced significant changes to academic research, transforming traditional methodologies and creating new possibilities for data analysis and interpretation, yet at the same time raising significant questions and concerns regarding the appropriateness and ethicality of their use. This study examines the impact of GenAI on research practices, focusing on its application in qualitative and quantitative data analyses. As these tools become more prevalent in the academic landscape, it is essential to examine the complex nature of their use in detail and aim to understand the ethical implications, frame the discussion, and avoid the uncritical adoption of technologies that are not yet fully understood.

While it may be argued that GenAI represents more 'style than substance' or is a part of a 'hype bubble' which will soon burst, we do not feel that this is the case. While these technologies have received a great deal of attention in the public sphere and are perhaps at times misunderstood or misestimated in their abilities, we nevertheless believe that the importance of GenAI in academic research is considerable. Tools such as ChatGPT and other Large Language Models (LLMs) have shown remarkable abilities to process large amounts of data, generate insights, and assist researchers across various disciplines (Ibrahim et al., 2023) while progressing and maturing in a staggeringly short period of time. As a result, their potential to improve research productivity, make complex analytical processes more accessible, and encourage innovative approaches to knowledge creation has attracted significant interest within the academic community (Kamalov et al., 2023). The use of GenAI tools in research also raises important questions regarding research integrity, authorship, and the changing nature of academic work (Cotton et al., 2023; Perkins & Roe, 2024a). As institutions place increasing emphasis on research output and rankings, the pressure on researchers to use these tools for increased productivity may unintentionally compromise fundamental principles of academic integrity or lead to strains in the systems of academic communication because of increased research outputs.

This study aims to provide an overview of the current and potential applications of GenAI in academic research, focusing specifically on qualitative and quantitative data analyses. By examining both the opportunities and challenges presented by these tools, we seek to contribute to the ongoing discussion of the responsible and ethical uses of GenAI in academia. To guide our exploration, we pose three key research questions.

1. How are GenAI tools currently being applied in qualitative and quantitative data analysis within academic research, and what potential future applications are emerging?
2. What are the key benefits and limitations of using GenAI tools in academic research, particularly in terms of research efficiency, accuracy, and generation of new insights?
3. What ethical considerations and challenges arise from the integration of GenAI tools in academic research and how can these be addressed to ensure research integrity and transparency?

Through a critical examination of these questions, we aim to provide insights that will assist researchers, institutions, and policymakers in effectively and ethically using GenAI-assisted research. We examine specific applications of GenAI in qualitative and quantitative research methodologies and present case studies and examples that illustrate both the potential and limitations of these tools. This includes exploring the use of GenAI in tasks such as transcription, coding, thematic analysis, visual analytics, and statistical analysis. We cover some of the technical limitations of GenAI tools, including issues of replicability, the 'black box' nature of some algorithms, and the challenges in interpreting complex, culturally mediated qualitative data. Furthermore, we explore the ethical implications of increased reliance on GenAI in academic research, considering issues such as researcher autonomy, potential biases, institutional pressures, and the need for transparency in AI-assisted studies.

By addressing these areas, we aim to contribute to the development of best practices for the integration of GenAI tools into academic research, balancing the potential benefits with the need to maintain the integrity and quality of scholarly work.

## Literature

The beginnings of GenAI in research and education rang alarm bells for many. At breakneck speed, tools to detect GenAI output were used to counter perceived risks to academic integrity. Later research demonstrated the insufficiencies of these detection tools (Perkins, Roe, Postma, et al., 2024; Perkins, Roe, Vu, et al., 2024; Weber-Wulff et al., 2023) and the ways in which they may penalise certain groups, such as non-native English speakers (Liang et al., 2023). At the same time, the analysis of university and publishing house policies demonstrated a movement away from the 'blocking' of GenAI tools in research to embrace their use within certain guidelines of transparency, honesty, and ethical use (Perkins & Roe, 2023). The narrative has now changed. While detection tools are still in use, a greater degree of nuance is required to interpret their output, and many researchers are already using GenAI to assist with writing and potentially with other forms of analysis and research (Perkins & Roe, 2024a). That said, it is still early days in the integration of GenAI into scholarly work. In a large scale Nature survey, 30% of researchers stated that they were using LLMs in research writing (Prillaman, 2024). Furthermore, it seems that the sentiment towards their use is generally optimistic; empirical studies have shown that researchers feel positive about the potential for using GenAI for research (Al-Zahrani, 2023; Geng & Trotta, 2024). At the same time, there is concern that these tools and their unique writing styles and propensity for producing certain phrases have already had a corrupting or homogenising effect on scientific publications, and a 2024 study showed that an estimated 35% of ArXiv abstracts demonstrated evidence of having been edited with ChatGPT (Geng & Trotta, 2024).

On the other hand, this high usage suggests that these tools serve a purpose and have utility for those who use them. Indeed, although there is significant hype around the use of GenAI and LLMs, especially in the media (Roe & Perkins, 2023), there is something to be said for taking a tool-based understanding of GenAI systems as computational instruments which can play a role in 'warm-blooded' research (Leslie, 2023). This role has been recognised by supranational and intergovernmental organisations, many of which seem to have accepted that intervention and guidance needs to be given to ensure that researchers are at least aware of good practice when using GenAI in research. The UNESCO (Miao & Holmes, 2023) and European Commission (2023) guidelines on the use of GenAI in research, for example, offer some insights and advice for good practices for individuals, as well as implications for institutions and funding bodies. While both sets of guidelines suggest that GenAI can be beneficial in the research process, they equally emphasise the importance of ethical and responsible AI use in research while highlighting its potential and risks.

Beginning with UNESCO (Miao & Holmes, 2023), the guidance outlines several potential applications of GenAI in research, such as developing research questions, suggesting methodologies, and automating the aspects of data interpretation. However, warnings of familiar risks include fabricated information, privacy breaches, and reinforcement of dominant social norms at the expense of diverse viewpoints. Both the UNESCO and European Commission guidelines emphasise the critical role of human involvement in the research process. They emphasised that researchers must possess strong subject knowledge to verify GenAI outputs and maintain the ability to critically evaluate content. The European Commission guidelines particularly emphasise researchers' ultimate responsibility for scientific output and the need for transparency in AI use. The discourse on GenAI use in academic publishing, media, and leading university academic conduct policies seems to match this perspective, as we have found in our earlier investigations (Perkins & Roe, 2023, 2024a; Roe & Perkins, 2023).

Furthermore, key recommendations from both sets of guidelines include providing guidance and training on GenAI for researchers, building capacity for effective prompt engineering, and developing institutional strategies for the responsible and ethical use of GenAI. Advice has also been provided to avoid the use of GenAI tools in sensitive activities, such as peer reviews or evaluations, although there has been mixed research on the potential benefits or problems that may arise in these instances (Checco et al., 2021; Zhou et al., 2023).

Drawing on these two publications as foundational literature before discussing more specific studies on the use of GenAI in qualitative and quantitative research, it is possible to set the scene regarding the general concerns and issues to be aware of for researchers seeking to make use of GenAI. This extends not only to issues of 'hallucinations' (an anthropomorphised term for the factually incorrect outputs of GenAI), but also the long-term impact on the homogenisation of research, copyright issues, and assessment of research quality in a world where GenAI enables supercharged publications and outputs for savvy researchers who are able to benefit from such productivity gains.

There is a seemingly symbiotic relationship between academia and GenAI, as interest from educators and institutions drives demand and new applications – the academic sector has become a rapidly expanding market for AI tools (Pinzolits, 2024). Some of the most notable benefits of GenAI in research include democratisation of the systems of scientific publication. (Barros et al., 2023), for example, note the potential to increase equity for researchers from the Global South and early career researchers (ECRs) by helping overcome language barriers, given that non-native English speakers are heavily disadvantaged when pursuing a career in science. (Amano et al., 2023). However, this trend may also lead other researchers (not necessarily from the Global South) to produce numerous, low-quality papers, leading to potential surge in inadequate or irrelevant manuscript submissions, named a 'coming tsunami' by (Tate et al., 2023). This was noted by Prillaman (2024), who stated that the impact of GenAI in research may be on those involved in the review and editorial process. However, where GenAI causes issues, solutions may not be far off. Solomon (2023) speculate that LLMs may have a place as an assistant in the peer-review process in future. On the other hand, drawing on a similar concern to the homogeneity of research described in the European Commission and UNESCO guidelines, authors such as Messeri and Crockett (2024) have argued that the use of GenAI may lead to the 'monoculturing' of scientific knowledge, while Watermeyer et al.. (2024) research suggests that the 'automation' of academia through AI may contribute to the dysfunctional aspects of neoliberal academia.

Despite such a burgeoning body of research on GenAI in education, research, and almost all other disciplines, surprisingly few published studies describe exactly how GenAI tools may be used in the analysis of data. Those which are available most often describe the use of GenAI applications, such as ChatGPT in qualitative research methodologies. Yan et al. (2024) conducted a user study involving ten qualitative researchers to explore the potential of ChatGPT as a collaborative tool in thematic analysis. Their findings demonstrated that the ChatGPT had value in enhancing coding efficiency, facilitating initial data exploration, providing granular quantitative insights, and aiding comprehension for non-native speakers and non-experts. On the other hand, the researchers also identified persistent issues regarding ChatGPT's trustworthiness, accuracy, reliability, and contextual understanding, as well as concerns about its broader acceptance within the research community. To address these challenges, Yan et al. (2024) proposed five design recommendations: implementing transparent explanatory mechanisms, improving interface and integration capabilities, enhancing contextual understanding and customisation, incorporating human-AI feedback loops and iterative functionality, and developing robust validation mechanisms to strengthen trust. However, whether such recommendations can be brought into reality given the current technological, economic, and scientific constraints is still unknown. Perkins and Roe (2024) described an experimental process using ChatGPT to support an inductive thematic analysis through triangulation, conducting separate human-driven and GenAI-assisted processes of coding and code development, and then using both datasets to arrive at a final set of themes; similarly, this process was impacted by hallucinations and irreproducibility in outputs, even given the same input data. The authors called for a high degree of criticality when attempting a similar method. Dahal (2024) postulates that GenAI tools may be helpful tools as co-authors and research assistants, while also addressing the potential ethical issues involved. Bijker et al. (2024) explored the utility of ChatGPT in conducting qualitative content analysis, analysing 537 forum posts about sugar consumption reduction using both inductive and deductive approaches. The study found ChatGPT to be fairly reliable in assisting with qualitative analysis, performing better with inductive coding schemes than deductive ones, and showing potential as a second coder with high agreement in some coding schemes.

Other authors have attempted to develop their own stand-alone applications to assist with qualitative data analysis. Gao et al., (2023) investigated AI-assisted collaborative qualitative analysis (CQA) by developing and evaluating a tool called CoAIcoder. Their research, involving 32 pairs of CQA-trained participants, explored four collaboration methods across common CQA phases and found that using a shared AI model as a mediator among coders could improve CQA efficiency and foster quicker agreement in the early coding stages, although this approach could affect the final code diversity. A further example is that of Gebreegziabher et al. (2023), who described PaTAT as an AI-enabled tool designed to support collaborative qualitative coding in thematic analysis. PaTAT employs an interactive program synthesis approach to learn flexible and expressive patterns from user-annotated code in real time. The tool addresses the challenges of ambiguity, uncertainty, and the iterative nature inherent in thematic analysis using user-interpretable patterns.

A further tool proposed by Hong et al. (2022), named Scholastic, is designed to support human-AI collaboration in interpretive text analysis using a 'machine-in-the-loop', human-centred approach. Scholastic incorporates a clustering algorithm to scaffold the analysis process in an attempt to address concerns regarding machine-assisted interpretive research. This system allows scholars to apply and refine codes, which then serve as metadata. This approach aims to enhance the scalability of interpretive text analysis while maintaining the integrity of human-driven interpretive research.

Leeson et al. (2019) conducted a proof-of-concept study to evaluate the potential of Natural Language Processing (NLP)for analysing qualitative data in public health research. The study compared two NLP methods, topic modelling and Word2Vec, with traditional open coding to analyse interview transcripts, and found that all three methods produced relatively similar results for most interview questions, with NLP methods being able to process large amounts of data rapidly. The authors suggest that NLP could serve as a useful adjunct to qualitative analysis, either as a post-coding check on accuracy or as a pre-coding tool to guide researchers. Another example is that of Lennon et al. (2021), who developed and tested an Automated Qualitative Assistant (AQUA) to support qualitative analysis in primary care research. When this tool was used with a large dataset of free-text survey responses, it was able to demonstrate intercoder reliability comparable to that of human coders in some areas.

Regarding attitudes towards the use of GenAI tools in the research process, Al-Zahrani (2023) surveyed university students (n = 505) on their perceptions of the use of GenAI tools and found mainly positive attitudes. However, given that the sample explored was not active HE researchers, this must be considered as a limitation to the study.

## GenAI Assisted Qualitative Data Analysis

As seen in the literature review above, GenAI tools have attracted attention for their natural language abilities in dealing with qualitative data. This has often been used to identify potential themes, content, or topics that recur within written texts, or to provide secondary assistance in a 'machine-in-the-loop' approach to analysis. Below, we describe in more detail some of these applications, their benefits, and potential drawbacks for use in certain research contexts.

**Transcription and Text Processing**

Transcription and text processing are some of the most immediate and impactful applications of GenAI in qualitative research. Advanced speech recognition algorithms combined with natural language processing capabilities have significantly improved the process of converting audio recordings into textual data. Platforms such as Microsoft Teams and Otter.ai (Otter.ai, n.d.) now offer real-time transcription services, allowing researchers to focus on the content of interviews or focus group discussions rather than on taking notes. GenAI tools can process these transcripts further, identify speakers, detect emotional tones, and suggest initial coding schemes based on their content (Perkins & Roe, 2024a). This capability not only saves time but also provides a preliminary layer of analysis that researchers can build on. For example, a GenAI tool could analyse a series of interview transcripts from a study on teacher burnout, highlighting recurring themes, emotional patterns, and potential areas for further investigation.

**Code Generation and Thematic Development**

The process of coding qualitative data, which is traditionally a time-intensive and subjective task, may be significantly enhanced by GenAI tools. These systems can rapidly analyse large volumes of textual data and identify patterns, recurring themes, and anomalies that might escape human observation (Jiang et al., 2021). Using natural language prompts, researchers can guide GenAI tools to generate initial coding schemes or refine existing ones (Bijker et al., 2024; Perkins & Roe, 2024a; Yan et al., 2024). Although these initial codes would require human validation and refinement, they could provide a starting point for further analysis.

GenAI tools also seem to excel in identifying latent themes, connections, and subtleties that may not be immediately apparent in human-centred analysis (Perkins & Roe, 2024a). By analysing the relationships between codes and the contexts in which they appear, these tools can suggest higher-order themes or theoretical constructs (Gao et al., 2023). This capability is particularly valuable in approaches such as grounded theory, where the goal is to develop theoretical insights from data (Sinha et al., 2024). However, it must also be accepted that the interpretation of certain forms of data, such as text transcripts, may not be deeply analysed by GenAI tools alone (Leeson et al., 2019). Semantic prosody, irony, sarcasm, emotion, paralinguistics, or other forms of non-verbal communication do not appear in text; thus, they are invisible to a GenAI tool. This is perhaps one of the biggest issues regarding accuracy and validity when using a GenAI tool to engage in qualitative data analysis (Hong et al., 2022) and highlights the fact that GenAI is best as a supportive or adjunct researcher with a specific set of very deep, yet not very broad skills.

**Case Study: Inductive Thematic Analysis with ChatGPT**

Our previous research, detailed in a paper in the Journal of Applied Learning and Teaching (Perkins & Roe, 2024b), demonstrated a novel approach to inductive thematic analysis using ChatGPT. This case study provides a practical example of how GenAI tools can be integrated into qualitative research methodologies. In this study, we employed a dual-analysis approach: one researcher conducted a traditional manual analysis, whereas the other utilised ChatGPT to assist in the analysis process. The dataset comprises policies related to the use of AI tools in academic research from various publishers' websites. GenAI-assisted analysis demonstrated several advantages. First, ChatGPT can rapidly generate an initial set of codes from the dataset, providing a solid foundation for further analysis. Additionally, the tool demonstrated proficiency in theme identification, suggesting potential themes based on the relationships between codes and offering new perspectives on data. Finally, the GenAI-assisted analysis proved to be more efficient than manual analysis, allowing researchers to dedicate more time to interpretation and theory development. However, this process presents certain challenges. The stochastic nature of ChatGPT led to inconsistencies, where repeated analyses of the same data using different versions of the tool yielded slightly different results. The quality and relevance of the AI output were also heavily dependent on the researcher's ability to craft effective prompts, highlighting the importance of prompt engineering skills in leveraging GenAI tools effectively. Despite these challenges, the integration of ChatGPT into the thematic analysis process has demonstrated significant potential for enhancing qualitative research methodologies. This highlights the importance of combining AI capabilities with human expertise to achieve robust and insightful analyses (Perkins & Roe, 2024b, 2024a; Sinha et al., 2024).

**Specific modes of analysis**

Building on a case study of inductive thematic analysis, GenAI tools show promise in various qualitative methodologies. In narrative analysis, which focuses on understanding and interpreting stories and their inherent meanings, these tools can identify recurring patterns, themes, and emotional trajectories within stories, potentially enhancing researchers' understanding of complex narratives (Jiang et al., 2021). For grounded theory approaches, which aim to develop theoretical explanations grounded in empirical data through an iterative process, GenAI can assist in rapid data categorisation, suggesting relationships between codes and categories, and even proposing theoretical constructs (Gebreegziabher et al., 2023). While narrative analysis is concerned with the structure and content of stories and grounded theory seeks to generate theory from data, despite these differences, both methodologies can benefit from the use of GenAI tools. However, although GenAI tools can augment these methodologies, they cannot yet replace human expertise (Perkins & Roe, 2024b). The researcher's critical thinking, contextual understanding, and interpretive skills remain essential for ensuring the validity and depth of qualitative analyses (Yan et al., 2024).

# Quantitative Data Analysis with GenAI

The integration of GenAI tools into quantitative research methodologies has the potential to revolutionise how researchers approach data analysis, interpretation, and visualisation, however there has been presently much less focus on the use of GenAI tools for quantitative analysis as opposed to qualitative analysis, partially due to concerns regarding the 'black-box' nature of these algorithms. However, recent advances in these tools have allowed for increased trust in their outputs, as discussed below. The rapid advancement of these tools is opening up new avenues for researchers to handle increasingly complex datasets and derive meaningful insights more efficiently than ever before (Kamalov et al., 2023), and here we provide examples of how GenAI tools may support in quantitative analysis.

**Visual Analytics and Pattern Identification**

One of the most significant contributions of GenAI tools to quantitative research is in the fields of visual analytics and pattern identification. GenAI applications may be able to rapidly process large datasets and generate sophisticated visualisations that help researchers identify trends, anomalies, and relationships that might not be immediately apparent through traditional statistical methods. The ability of GenAI tools to generate these visualisations through natural language prompts makes them particularly powerful. For instance, a researcher could upload a dataset and simply ask, 'Show me the correlation between various economic indicators and stock market performance over the past decade'," and the tool would generate the appropriate visualisation, potentially revealing insights that might have been missed through traditional analysis methods. This capability not only saves time but also allows researchers to quickly explore data from multiple angles, potentially leading to new hypotheses and research directions.

Developments in existing GenAI tools, such as Claude and ChatGPT, allow for increased data visualisation possibilities. As an example, in 2024, Anthropic added a feature named 'Artifacts' to Claude, allowing it to not only interpret uploaded data, but also produce visual artifacts to aid in data presentation. For example, it can create custom charts, graphs, and even infographics based on complex datasets, making it easier for researchers to communicate their findings effectively (Anthropic, 2024). These advanced capabilities are particularly useful in fields dealing with big data, such as genomics, climate science, and social network analysis, where traditional visualisation methods may struggle to capture the full complexity of the data. However, while these tools can generate complex visualisations, the interpretation and contextualisation of these visual outputs still require human expertise and domain knowledge to ensure they accurately represent the data and align with the research objectives.

**Integration with Statistical Software**

GenAI tools are increasingly being integrated with popular statistical software and libraries to create hybrid systems that combine the strengths of traditional statistical methods with AI-driven analytics. Platforms such as Python and R, which are widely used in academic research, are now integrated into GenAI tools (Machlis, 2023), allowing researchers to leverage a broader range of analytical capabilities. For example, a researcher working on a complex longitudinal study could use a GenAI tool to preprocess and clean the data, identify potential outliers or missing data patterns, and suggest appropriate statistical models based on the data characteristics. The researcher could then use traditional statistical software to run the analyses with the GenAI tool assisting in the interpretation of results and generation of visualisations. This synergy between AI and traditional statistical methods can lead to more robust and comprehensive analyses, particularly when dealing with large, complex datasets and may help to bridge the gap between AI capabilities and established research methodologies, making it easier for researchers to adopt these new technologies without completely overhauling their existing workflows, and at the same time, reducing concerns related to reproducibility of results

**Natural Language Interactions**

Natural language interactions for complex analyses represent another significant advancement introduced by GenAI tools in quantitative research. These tools are becoming increasingly capable of interpreting natural language queries and translating them into appropriate statistical procedures. This capability has the potential to democratise advanced statistical analyses, making them accessible to researchers who may not have extensive training in statistical methods. For example, a researcher could ask a GenAI tool to conduct a multiple regression analysis on selected variables and explain the results, or carry out analysis in traditional statistical software, and then ask for an interpretation by a GenAI tool. While this capability is powerful, researchers must develop and maintain a solid understanding of statistical principles to critically evaluate AI-generated analyses and ensure their appropriateness for the research questions at hand. The democratisation of advanced statistical techniques through language interfaces could lead to more interdisciplinary research and collaboration, as researchers from diverse backgrounds can more easily engage with complex quantitative methods. However, this also raises concerns about the potential for misuse or misinterpretation of statistical results by users who may not fully understand the underlying principles, highlighting the need for continued education and training in statistical literacy alongside the adoption of these AI tools or their further integration into existing software tools.

## Limitations and Ethical Considerations of GenAI Tools in the Research Process

**Technical Limitations**

Despite the potential use of GenAI tools in research, several limitations must be acknowledged. One significant concern is the potential of these tools to produce statistically significant but spurious correlations, particularly when dealing with large datasets. The ease with which GenAI tools can generate analyses and visualisations may lead to increased cases of "p-hacking", where researchers inadvertently or intentionally search for patterns that lack theoretical significance. Given that this is an extant problem in academic research (Head et al., 2015), changes in behavioural patterns to make this simpler may be of concern when considering research integrity.

Another consideration is the "black box" nature of some GenAI algorithms, which can make it difficult to fully understand and explain the process by which certain results were obtained. This lack of transparency can be problematic in academic research, where replicability and clear methodologies are paramount (Nichols et al., 2021). Furthermore, the stochastic nature of many GenAI tools can lead to inconsistencies in the results, even when using the same data and prompts. This variability poses challenges to the replicability of research findings (Perkins & Roe, 2024a), a cornerstone of scientific enquiry. For instance, in our case study using ChatGPT for thematic analysis (Perkins & Roe, 2024b), we observed that repeated analyses of the same dataset yielded slightly different themes or coding structures. This inconsistency raises questions about the reliability of GenAI-assisted analyses and the extent to which research findings can be replicated.

When dealing with qualitative data, especially if for example, data is captured through ethnographic, interview, or observational methods, a great deal of the 'substance' in the studied phenomenon cannot be understood merely through texts or through images. Subjective experience, empathy, and deep interpretation of context require the skills that AI applications currently struggle with, especially considering the culturally biased nature of these tools (Spennemann, 2024). Therefore, this significantly affects and potentially impacts our ability to view studied phenomena and data outside of the lens of a Western perspective on the world (Roe, 2024) and is a serious consequence of using a culturally oriented tool for a specific research process without critically evaluating its purpose.

A further problem comes from language – while GenAI applications may be able to effectively identify recurring themes, topics, and structures in datasets – which can be important and relevant – such tools cannot hope to interpret the meaning of a pause, the thinking behind an extended period of silence, or the true

'meaning' or significance of a lived experience. On the other hand, this does not mean that GenAI is of no use even when exploring such datasets; in fact, even in ethnography, AI-driven methods are being proposed by leading anthropology organisations to enhance the research process (Artz, 2023).

**Ethical Considerations**

The integration of GenAI tools in academic research raises significant questions regarding researcher autonomy, attribution, and research integrity. While these tools can enhance efficiency and provide new analytical capabilities, there is a risk of over-reliance on AI, potentially diminishing the researcher's role in shaping research direction and interpretation (Cotton et al., 2023; Perkins, 2023). Researchers must maintain their critical thinking skills and domain expertise using GenAI tools as aids, rather than replacements for human judgment (Bearman & Ajjawi, 2023). The interpretation of results, development of theoretical frameworks, and drawing meaningful conclusions should, therefore, remain firmly in human hands.

Although GenAI tools can increase efficiency in certain aspects of research, they should not be seen as a replacement for the critical thinking and domain expertise of researchers. Human supervision, monitoring, and expert input are essential, particularly when novice researchers use GenAI to assist in research they do not have a deep understanding of. Although a 'human in the loop' may eventually be possible for some forms of analysis, a 'machine in the loop' approach, where GenAI supports rather than replaces human researchers, is recommended to leverage the benefits of these tools while maintaining research integrity (Bearman & Ajjawi, 2023; Perkins & Roe, 2024b), and where the time saved with the use of these tools should instead be redirected towards deeper analysis, interpretation, and theoretical development. In other words, the present abilities of GenAI applications and their associated limitations require supervision, monitoring, and additional input from expert human researchers familiar with the subject matter. In cases where a novice researcher seeks to use GenAI tools to assist in researching a new subject, it is critical to carefully evaluate the credibility and accuracy of the information provided by GenAI because the consequences of relying on false or misleading information can be severe. However, without a deep understanding of the subject, this may not be possible.

The potential for GenAI tools to increase research productivity coupled with institutional pressures to improve research output and rankings could create problematic incentives that prioritise quantity over quality (Miao & Holmes, 2023). The academic community should develop guidelines for responsible GenAI use, emphasising novel insights and theoretical contributions over output quantity. This includes the challenges of authorship and attribution when AI generates significant portions of the research content. Clear guidelines must be developed to ensure proper credit for human intellectual contributions while acknowledging AI involvement (Perkins & Roe, 2024b).

Resnik and Hosseini (2024) highlight additional ethical considerations, including the need to identify and control AI-related biases, engage with impacted communities, and properly handle synthetic data. They emphasise that while AI use does not necessitate changing the established ethical norms of science, it requires new guidance for appropriate use. A critical challenge is the potential for GenAI tools to replicate and amplify existing biases, even when they are not apparent in the datasets (Hacker et al., 2024). The interpretation of AI-generated results is key, as human researchers may inadvertently reinforce societal biases through their analyses and conclusions. This underscores the importance of diverse research teams and ongoing critical examination of AI-assisted research processes and outcomes.

These questions require open dialogue and discussions within the academic community. Clear guidelines must be developed to ensure transparency and maintain the integrity of the academic publishing process. This could include requiring detailed methodological sections that describe AI involvement or even the creation of AI disclosure statements similar to conflict of interest disclosures (Crawford et al., 2023). For example, in our case study using ChatGPT for thematic analysis (Perkins & Roe, 2024b), we explicitly described the process of using the AI tool, including the prompts used and challenges encountered. This level of transparency allows other researchers to critically evaluate the findings and replicate the study. Furthermore, transparency extends to acknowledging the limitations of the GenAI tools in the research process. Researchers should be aware of the potential for AI-generated errors or biases and describe the steps taken to validate and verify the AI-generated results. This might include cross-checking AI outputs against manual analyses or using multiple AI tools to triangulate findings. Additionally, researchers should consider using GenAI tools in combination with traditional methods, using AI-generated insights as a starting point for further human analysis and interpretation.

## Future Considerations

As GenAI technology continues to advance rapidly, we anticipate several developments that will further impact academic research. Future iterations of GenAI tools are likely to offer improved accuracy and consistency in their outputs, potentially addressing some of the current concerns regarding replicability (Perkins & Roe, 2024b). As several academic publishers have already announced licencing agreements with GenAI tool developers (Dutton, 2024), we foresee the development of more specialised GenAI models specifically trained on academic literature and research methodologies, potentially offering more nuanced and context-aware assistance in research tasks. Advancements in explainable AI could lead to GenAI tools that provide clearer insights into their decision-making processes, addressing some of the current "black box" concerns. Furthermore, GenAI tools may become more seamlessly integrated into existing research software and workflows, making their use more intuitive and efficient. It is possible that AI systems will be able to provide deeper, more humanistic analysis – including of nuanced emotional states, multicultural worldviews, and more sophisticated interpretations of the context, symbolism, and meaning of invisible data – although developments of this nature will equally raise further questions.

The increasing capabilities of GenAI tools may lead to significant shifts in how academic research is conducted. We may see the emergence of new research methodologies that blend traditional approaches with AI-assisted techniques, potentially bridging the gap between qualitative and quantitative methods. The speed of GenAI tools could enable more real-time analysis of data, potentially allowing for more dynamic and adaptive research designs, which could include future research teams including AI "collaborators" working alongside human researchers, each contributing their unique strengths to the research process. This collaboration could lead to new forms of knowledge creation and dissemination, challenging traditional notions of authorship and research output, and meaning publishers may need to consider their present attitude towards GenAI tools as authors (Perkins & Roe, 2024a). At the same time, new challenges to research integrity and security will also emerge, as GenAI-driven technologies such as deepfakes begin to make it more complex to secure data using traditional methods of authentication (Roe & Perkins, 2024). Additionally, ensuring the privacy and ethical use of research data will become increasingly important as GenAI tools become more adept at processing and analysing large datasets. Continued attention will need to be paid to identifying and mitigating biases in GenAI tools, particularly when these tools are used in research that impacts vulnerable populations. This may result in the development of new standards and practices to ensure the integrity of AI-assisted research, including methods for verifying AI-generated results and ensuring transparency in AI use. These ethical considerations will need to evolve alongside the technology, requiring ongoing dialogue and collaboration between researchers, ethicists, and AI developers.

## Conclusion

This study explored the transformative impact of GenAI tools on academic research by examining their applications in both qualitative and quantitative data analyses. We have highlighted how these tools can enhance research efficiency, uncover new insights from data, democratize complex analyses, and potentially lead to new insights and methodologies. However, we have also identified significant challenges, including concerns about replicability, the importance of prompt engineering, and the ethical implications of increased AI integration in research.

Our analysis reveals that while GenAI tools offer significant potential to enhance research methodologies, from automating initial coding processes to facilitating complex statistical analyses, their use raises important questions about research integrity, authorship, and the changing nature of academic work. It is clear that while GenAI can increase research productivity, it should not replace the critical thinking and domain expertise of human researchers. There is a pressing need for transparency in AI-assisted research and clear guidelines on how to attribute AI contributions. We have highlighted several key applications of GenAI in research, including transcription and text processing, code generation, thematic development in qualitative analysis, visual analytics, integration with statistical software, and natural language interactions in quantitative analysis. However, significant technical limitations and ethical considerations must be addressed. These include the potential for spurious correlations, the "black box" nature of some GenAI algorithms, challenges to replicability, and the inability of current AI systems to fully capture the nuances of qualitative data, especially in contexts requiring deep cultural understanding or interpretation of non-verbal cues. Ethically, the risk of over-reliance on AI, potential biases in AI-generated results, and challenges to research integrity and authorship must be carefully navigated.

Looking to the future, we anticipate continued rapid advancements in GenAI technology, potentially leading to more accurate, consistent, and context-aware research tools. However, these developments will likely bring new ethical challenges and questions about the nature of academic research and authorship. In conclusion, while GenAI tools offer exciting possibilities for enhancing academic research, their responsible and ethical use requires ongoing attention, discussion, and adaptation within the academic community. As these tools continue to evolve, so too must our approaches to fully leveraging their capabilities while maintaining the integrity and quality of academic research. This is a challenging balancing act that will continue to develop in line with changing societal and academic norms regarding the use of GenAI tools for research, as well as the continued development of the technology itself.

**AI Usage Disclaimer**

This study used Generative AI tools (Claude 3.5 Sonnet) for content development, revision and editorial purposes throughout the production of the manuscript. The authors reviewed, edited, and take responsibility for all outputs of the tools used in this study